\documentclass[a4paper,11pt]{article}
\usepackage{preamble}

\title{The spectrum of GUT-like gauge-scalar models} 
\author*{Elizabeth Dobson}
\author{Axel Maas}
\author{Bernd Riederer}

\affiliation{Institute of Physics, NAWI Graz, University of Graz,\\
  Universitätsplatz 5, Graz, Austria}

\emailAdd{elizabeth.dobson@uni-graz.at}
\emailAdd{axel.maas@uni-graz.at}
\emailAdd{bernd.riederer@uni-graz.at}

\abstract{
    Past lattice simulations tentatively suggested that the spectrum of observable particles in BSM theories is qualitatively different than perturbatively expected. 
	The discrepancy can be traced back to nontrivial field-theoretical effects arising from the requirement of gauge invariance.
	A perturbative but gauge-invariant approach working directly with bound states (as proposed by Fröhlich, Morchio and Strocchi) could provide a solution, by allowing composite-state effects to be treated within a perturbative framework. We consider a toy GUT-like theory -- $\SU(3)$ Yang--Mills coupled to a scalar `Higgs' in the fundamental representation -- and expand on previous work by providing the most comprehensive spectroscopy to date, including all channels up to spin 2.
	Our results strongly support the general conclusion that the elementary spectrum is not an adequate proxy for the low-energy spectrum of a GUT, and also suggest a possible analytical approach to BSM model-building via the Fröhlich--Morchio--Strocchi approach.
}

\FullConference{%
  The 39th International Symposium on Lattice Field Theory (Lattice2022),\\
  8-13 August, 2022 \\
  Bonn, Germany 
}

\begin{document}
\maketitle

\section{Introduction}
Grand Unified Theories (GUTs) are an appealing class of BSM theories due to their relatively minimal theoretical assumptions and their elegant explanation of both quantization of electric charges and of anomaly cancellation in the Standard Model (SM) \cite{Langacker:1980js}.
A key question of GUT research is: given experimental data describing physics at different energy scales, how can the Standard Model be embedded into a larger gauge group, such that the low-energy effective theory is consistent with the observed data? The answer to this is more complex than it initially appears.
\subsection{The importance of systematic control for GUTs}
For the sake of simplicity, consider a pure gauge-scalar theory (i.e. no fermions) with some gauge group $\mathcal{G}$.
Perturbation theory typically assumes not only that the gauge coupling is weak, but \emph{also} that the Higgs mechanism applies in the same way as for the Standard Model case --- that is, that the effective theory can be treated as having the gauge group $\mathcal{H}\subseteq\mathcal{G}$, where the freeze-out of degrees of freedom at low energies can be interpreted as the `spontaneous breaking' of $\mathcal{G}$ down to $\mathcal{H}$.
This approach is of course questionable from a nonperturbative viewpoint: by Elitzur's theorem \cite{Elitzur:1975im}, `spontaneous' breaking of gauge symmetries is only possible as a consequence of fixing to a certain gauge.
A priori, there is no reason to assume that a certain subgroup $\mathcal{H}$ of $\mathcal{G}$ should coincide with the low-energy limit\footnote{The situation becomes more subtle when $\mathcal{G}$ has multiple little groups, which could in principle allow for different `choices' of breaking patterns: see \cite{Dobson:2022ngz} for a more detailed discussion.}, and while this works for the Higgs mechanism in the SM context, this cannot be naively generalised to GUTs.
Moreover, the perturbative state space of elementary fields is unphysical. This is due to the Gribov–Singer ambiguity, which invalidates the perturbative Becchi--Rouet--Stora--Tyutin (BRST) construction as a means to identify the physical degrees of freedom in the nonabelian case, even at arbitrarily weak coupling \cite{Gribov:1977wm,Singer:1978dk,Fujikawa:1982ss}.
\footnote{In brief, the Gribov--Singer ambiguity states that, for a nonabelian gauge theory, it is impossible to obtain a continuous choice of unique representative on each gauge orbit from a local gauge condition alone. 
This means that there are distinct transverse configurations satisfying any gauge condition, related via large gauge transformations, which together define the \emph{residual gauge orbit} \cite{Maas:2017csm}.
Thus nonabelian gauge theories impose additional conditions compared to abelian ones, as a supplemental gauge condition is required to physically identify the distinct Gribov copies.
%The actual physical configuration space is restricted to the \emph{fundamental modular region}, the region of configuration space free of Gribov copies, which is topologically nontrivial for nonabelian gauge theories.
}

All of this, and the necessity to perform calculations in arbitrary GUTs, motivates the need for a perturbative description with a physical state space.
Such a construction, as described by Fröhlich, Morchio and Strocchi (FMS) \cite{Frohlich:1980gj,Frohlich:1981yi} attempts to resolve the problem of gauge invariance by working with composite fields directly.
We will refer to their approach from here on as `(FMS-)augmented perturbation theory'.
The basic principle is to provide a mapping between such observables with respect to an ``unbroken'' gauge group $\mathcal{G}$, and their counterparts in the ``broken'' theory described by the gauge group $\mathcal{H}$ via a double expansion in first the fluctuation over the vacuum expectation value (vev), $(\eta/v)$, and second in the gauge coupling.
Choosing a gauge with nonzero vev hides the symmetry, but leaves it intact. Rewriting the scalar field in terms of this gauge provides an exact, nonperturbative equation, where the LHS corresponds to the $\mathcal{G}$-observable, and the RHS to a sum of observables of the $\mathcal{H}$ theory.
Each term on the RHS is $\mathcal{G}$-dependent, but their sum is $\mathcal{G}$-invariant. 
In this way we can create an exact, nonperturbative mapping between bound states  and the gauge-dependent elementary states. 
As we discuss below, this mapping is in general nontrivial, which can lead to qualitative discrepancies between naive and FMS-augmented perturbation theory.
The hope of all of this, is that we can keep the convenience of perturbation theory whilst evading the issues of the Gribov-Singer ambiguity.

\subsection{The Fröhlich--Morchio--Strocchi construction: mapping elementary and composite states}
The `physical' version of the Higgs particle for an $\SU(N)$ theory with a single scalar $\phi(x)$ is the composite operator $(\phi^\dag\phi)(x)$ \cite{Frohlich:1980gj,Maas:2017xzh}. 
Substituting the splitting $\phi(x) = \tfrac{v}{\sqrt{2}}\hat{n}+\eta(x)$ with vev $v$ and fluctuation $\eta$ in a suitable gauge, the gauge-invariant Higgs propagator can be written as
\begin{equation}\label{eq:higgsprop}
  \Ev*{(\phi^\dag\phi)(x)(\phi^\dag\phi)(y)}\conn
  = v^2\Ev*{h(x)h(y)}\conn
  + 2v\Ev*{h(x)(\eta^\dag\eta)(y)}\conn
  + \Ev*{(\eta^\dag\eta)(x)(\eta^\dag\eta)(y)}\conn\,,
\end{equation}
where $h(x) = \sqrt{2}\Re[\hat{n}^\dag\eta(x)]$ is the usual elementary (unphysical) Higgs. 
The terms containing $\eta$ fields, neglected in the naive perturbative approach, are necessary to ensure gauge invariance. 
However, to leading order in perturbation theory, all that remains is the elementary Higgs propagator and a scattering state of two Higgs particles \cite{Maas:2017wzi}.
It can be shown for the Standard Model -- and it also seems possible for general $\SU(N)$ gauge theories -- that the poles of the propagator are contained within the 
% first term on the RHS of \cref{eq:higgsprop} 
leading order term in the vev-expansion
at all orders in perturbation theory \cite{Maas:2020kda}.
Therefore, even though the additional terms are technically required for gauge invariance of the propagator and to ensure the positivity of the Källén–Lehmann spectral representation, the effect is minimal: for this special case, the `physical Higgs' $(\phi^\dag\phi)$ can be well-approximated by its elementary counterpart $h(x)$. 

In general, however, the correspondence between $\mathcal{G}$-singlets and $\mathcal{H}$-singlets under gauge fixing is less straightforward.
The Higgs propagator in \cref{eq:higgsprop} belongs to a class of operators that can be uniquely constructed by considering a $\mathcal{G}$-observable at a fixed order in the FMS expansion in $(\eta/v)$ for the above gauge choice, without further multiplet decompositions, and it is further special in that it corresponds to the leading-order term in the fluctuation expansion.
For operators which do require an additional multiplet decomposition, or which are are split across different orders of the FMS expansion, the mapping between bound and elementary states is nontrivial \cite{Sondenheimer:2019idq,Maas:2017xzh}.
There are further possibilities \cite{Sondenheimer:2019idq}: the $\mathcal{H}$-singlets could be charged under a remnant abelian subgroup of $\mathcal{G}$ after gauge fixing; there could be $\mathcal{H}$-invariant observables with no $\mathcal{G}$-invariant counterparts; or even the converse, $\mathcal{G}$-observables with no $\mathcal{H}$-singlet counterparts. 
In general such situations are hard to calculate explicitly via perturbation theory, and evidence of the last of these could only be obtained via  non-perturbative calculations, e.g. observing states via lattice simulations which could not otherwise be accounted for.

\subsection{The spectrum of \tp{$\SU(N)$}{SU(N)} Yang--Mills--Higgs theory}

Fröhlich, Morchio and Strocchi \cite{Frohlich:1981yi} describe how other observables correspond to composite states in a similar way to \cref{eq:higgsprop}.
For example, considering the electroweak sector of the SM for simplicity, the `physical' version of the electron and neutrino is the scalar-fermion bound state\footnote{A consequence is that the Yukawa interaction has the natural interpretation of a dynamical oscillation phenomenon, without requiring a nonzero scalar vev \cite{Egger:2017tkd}.} $\phi^\dag\psi$.

In an SU($N>$2) theory, an analogue of the elementary vector boson $W_\mu$ is $i \phi^\dag D_\mu \phi$. The latter is instructive in understanding how the gauge-dependent viewpoint can lead to wrong predictions. To leading order, the physical version of the vector boson propagator is
\begin{equation}\label{eq:vectorprop2}
	\Ev{(i\phi^\dag D_\mu\phi)(x) (i\phi^\dag D_\nu\phi)(y)}\conn = v^2 \Ev{W_\mu^{(N^2-1)}(x) W_\nu^{(N^2-1)}(y)}\conn + \ord{(\eta/v)} + \dots\,,
\end{equation}
so the bound state is mapped to the heaviest gauge boson. For other investigated channels \cite{Maas:2017xzh,Sondenheimer:2019idq} no two-point function on the right-hand side has been found --- in fact, one does not expect that degenerate states exist in the physical spectrum \cite{Sondenheimer:2019idq}. This led to the prediction of the theory being gapped \cite{Maas:2016ngo,Maas:2017xzh}.
Thus, in contrast to \cref{eq:higgsprop}, this is already a situation where the FMS approach gives qualitatively different predictions to naive perturbation theory.
\footnote{At least, for $\SU(N>2)$. For the SM case, which is substantially different and no discrepancies appear at this order, we refer to \cite{Maas:2017wzi,Maas:2020kda}.}
Compare this to what one would expect if the Brout-Englert-Higgs (BEH) effect were to `spontaneously' break the $\SU(N)$ gauge group down to $\SU(N-1)$.
The fundamental spectrum, in this scenario, would consist of a single massive `Higgs' field, $(N^2-2N)$ massless gauge bosons, and $(2N-1)$ massive gauge bosons, with the lightest $(2N-2)$ degenerate. All other channels would contain only scattering states.
Note also that a theory with a nontrivial residual gauge group would, from a gauge-invariant perspective, imply the possibility of the presence of charged bound states which are absent from the perturbative spectrum.

Together this motivates a comparison of the gauge-dependent and gauge-invariant spectra, with the longer-term aims of identifying the phase structure and lines of constant physics for theories of this type.
So, starting from a toy model, we measure the physical spectrum via lattice simulations and compare it to the elementary spectrum. This allows us to assess the consequences of enforcing gauge invariance.
\section{Lattice tests of the FMS construction}
Lattice simulations allow us to test the predictions of the FMS mechanism nonperturbatively, and also give tentative hints about the kinds of features that a `GUT-like' model might exhibit.
Since simulating a realistic GUT is not computationally feasible, we work with a minimal toy model: here $\SU(3)$ Yang--Mills--Higgs theory with a single scalar in the fundamental representation, given by the action
\begin{multline}
  S[\phi,U] = \sum_x\Big\{
    (\phi^\dag\phi)(x) + \gamma\big[(\phi^\dag\phi)(x)-1\big]^2
    + \frac{\beta}{N}\Re\tr\sum_{\mu<\nu}\big[\one-U_{\mu\nu}(x)\big]
    -\kappa\sum_{\pm\mu}\phi^\dag(x) U_\mu(x)\phi(x+\emu)
    \,,
  \Big\}
\end{multline}
where $U_\mu$ are the link variables and $U_{\mu\nu}$ are the plaquettes, see \cite{Montvay:1994cy} for more details. 

Despite its artificial simplicity, this model has several useful properties.
The potential allows for a BEH effect at tree level, and yet
the nonabelian gauge group implies the failure of the BRST construction, which makes the model appropriate for testing the predictions of the FMS approach.
In addition, there is a nontrivial $\U(1)$ global group under which the scalars carry a charge of $\tfrac{1}{3}$.
In contrast to the SM, the global group and the gauge group are not of the same type, and the global group is smaller than the gauge group. It is this feature, which is common to most GUTs, which appears to be decisive in the emerging mismatch between the elementary spectrum and the physical spectrum \cite{Maas:2017wzi,Maas:2017xzh,Sondenheimer:2019idq}.
Investigating this model therefore allows the study of generic features that arise with larger gauge groups than that of the SM. 

In addition, predictions from FMS-augmented perturbation theory are available for some channels of this theory 
\cite{Maas:2016ngo,Maas:2017xzh}.
It is expected that if the FMS construction is a good approximation for $\SU(N)$ gauge-scalar theory with $N=3$, then it will also work for arbitrary $N>2$.
\Cref{sec:results} shows that this model is at the very least a suitable proof-of-concept to demonstrate the risks of overgeneralising from the SM case.

Previous tests of this model at different couplings $(\beta, \kappa, \gamma)$ indicated the existence of at least one custodially-charged operator for the fundamental $\SU(N)$ gauge-scalar case, and suggested the presence of distinct `Higgs-like' and `QCD-like' phases.\footnote{Note that it is currently unclear whether this corresponds to a `true' phase transition or just a crossover; while gauge invariance forbids a physical phase transition for the Standard Model/$\SU(2)$ case \cite{Osterwalder:1977pc}, a richer phase structure is possible for larger gauge groups.}
We extend this to a more detailed multi-channel investigation in the fundamental case and aim to determine the low-lying spectrum and its dependence on the couplings.
\subsection{Gauge-invariant operators}
To probe the FMS mechanism, we create a large basis of interpolators for gauge-scalar bound states analogous to the mesons and baryons of QCD, following a more basic version of the method outlined in, e.g. \cite{Dudek:2010wm}.
We consider gauge-invariant `building block' operators (chosen to have minimal field content in order to reduce noise) of the form \cite{Dobson:2021sgl,Dobson:2023}
\begin{align}
    \label{eq:om}
    f_{\mu_1,\dots,\mu_N} &= \phi^\dag D_{\mu_N}\dots D_{\mu_1}\phi\\
    g_{\mu_1,\dots,\mu_N; \{r,s\}} &= \epsilon_{abc}(D_{\mu_N}\dots D_{\mu_{s+1}}\phi)_a (D_{\mu_s}\dots D_{\mu_{r+1}}\phi)_b(D_{\mu_r}\dots D_{\mu_1}\phi)_c\,,
    \label{eq:ob}
\end{align}
where latin indices denote gauge indices.
These operators correspond to the custodially-neutral and charged cases respectively.
Other operator types are standard glueballs and pure-scalar states like $(\phi^\dag\phi)(x)$.
Currently, all operators are considered at zero-momentum, averaged over timeslices, and we construct linear combinations of either $f$-type or $g$-type operators that project simultaneously onto a definite continuum spin $J$ (up to $\ord(a^2)$ discretisation effects) and a specific irreducible representation $\Lambda$ of the cubic group.
Only after both projections are the operators discretised.\footnote{For these purposes, a non-improved lattice derivative $(D_\mu\chi)^\text{lat}(x) = \frac{1}{2a}\bq*{U_\mu(x) \chi(x+a\hat{\mu}) - U^\dag_\mu(x-a\hat{\mu})\chi(x-a\hat{\mu})}$ (where $\chi$ is any scalar lattice function) is sufficient to avoid discretisation errors}.

In this way, we create a basis of operators across all neutral, charged, and custodially-charged $J^P$ quantum number channels, subcategorised by their irreducible representation on the lattice.
In principle, operators with arbitrary spins and/or higher custodial charges can be implemented using the same methods, but due to statistical noise, we have currently limited our search to $J\leq 2$ and only a single charge under the $\U(1)$ symmetry.
The basis was enlarged by applying smearing (stout \cite{Morningstar2003gk}  and APE\cite{Falcioni:1984ei}) to the fields up to five times, and together with the enhanced variational analysis and fitting techniques of \cite{Jenny:2022atm} it was possible to detect ground-state signals in most channels (see \cref{fig:spectrum} below).

\subsection{Implementation details}
To perform the spectroscopy, we ran simulations on an $L^4$ isotropic lattice with $L=10,12,\dots,32$,
and used the results of \cite{Maas:2018xxu} to identify useful simulation points in terms of the couplings $(\beta,\kappa,\lambda)$.
The fields were updated using heatbath and overrelaxation algorithms. 
For the links, exact updates were possible using a modified version of the Cabbibo-Marinari method \cite[§7.8]{degrand2006}.
For the scalars, we used an approximate heatbath as described in \cite[A.C]{Knechtli:1999tw}, generating proposals according to a Gaussian distribution and accepting them according to the remaining quartic term of the potential, with the quadratic-quartic splitting chosen to maximise the acceptance probability. 
Further details of the implementation can be found in \cite{Dobson:2022ngz}.
% The fields were updated using the standard heatbath and overrelaxation algorithms: i.e. using a modified version of the Cabbibo--Marinari method to update the fundamental links exactly
% \cite[§7.8]{degrand2006}, together with an approximate scalar heatbath as described in \cite[A.C]{Knechtli:1999tw}, where proposals are generated according to a Gaussian and accepted according to the remaining quartic term of the potential, with the quadratic-quartic splitting chosen to maximise the acceptance probability.
% Further details of the implementation can be found in \cite{Dobson:2022ngz}.

We present here results for two cases, one in the region where the BEH effect occurs in a gauge-fixed setup and the (gauge) coupling is weak, and one where this is not the case. This theory, as is typical for scalar-gauge theories, has a low signal-to-noise ratio, so our results are statistics-limited.
\section{Features of the fundamental spectrum}
\label{sec:results}
\begin{figure}
    \centering
    \hspace*{-.8cm}
    \begin{subfigure}{.52\textwidth}
    \centering
    \includegraphics[height=5.75cm]{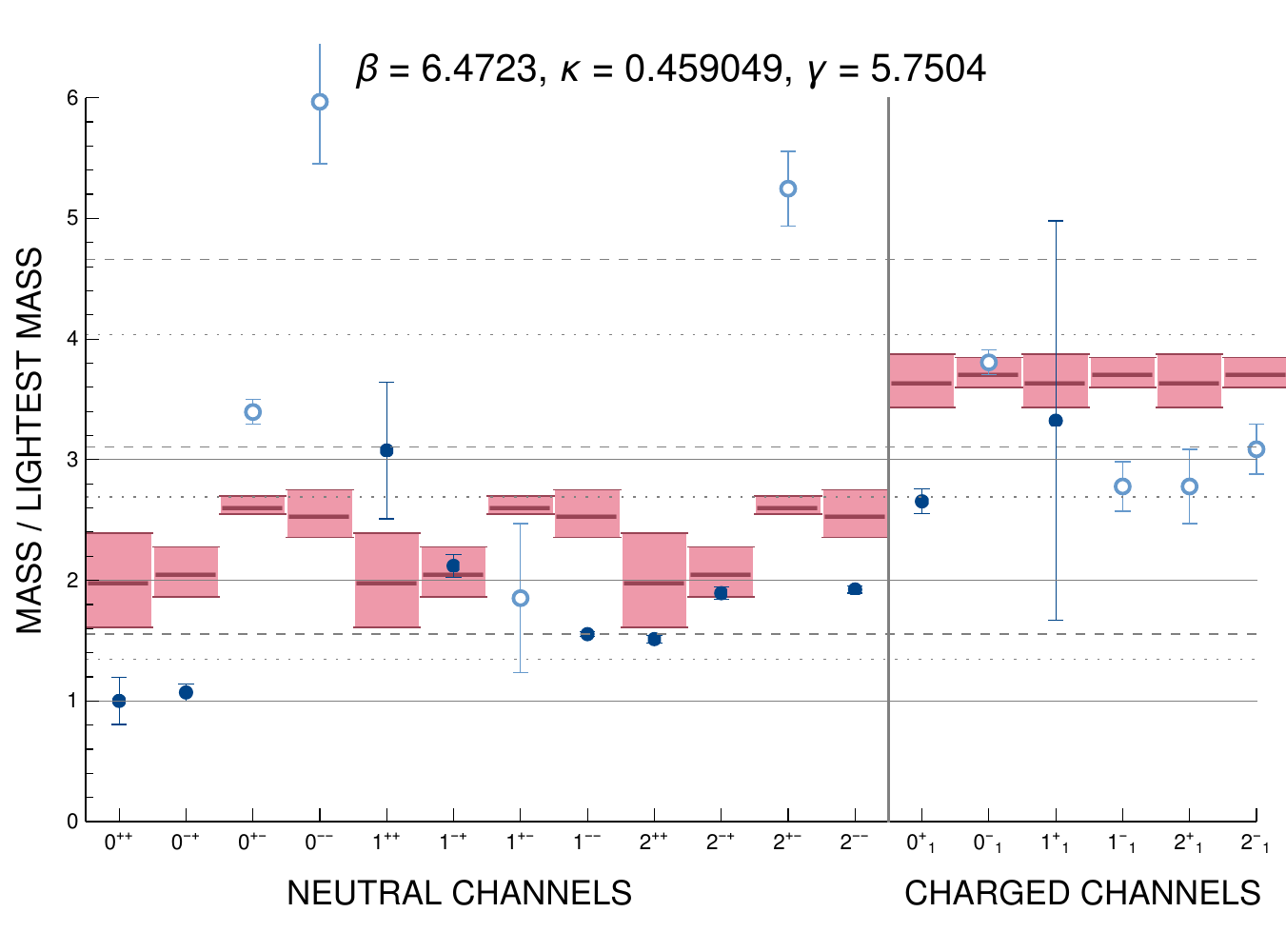}
    \vspace*{-.4cm}
    \caption{Without BEH effect, in the bulk `QCD-like' phase}
      \label{fig:sub:a}
    \end{subfigure}
    \begin{subfigure}{.5\textwidth}
    \centering
    \includegraphics[trim= 15 0 0 0,clip,height=5.75cm]{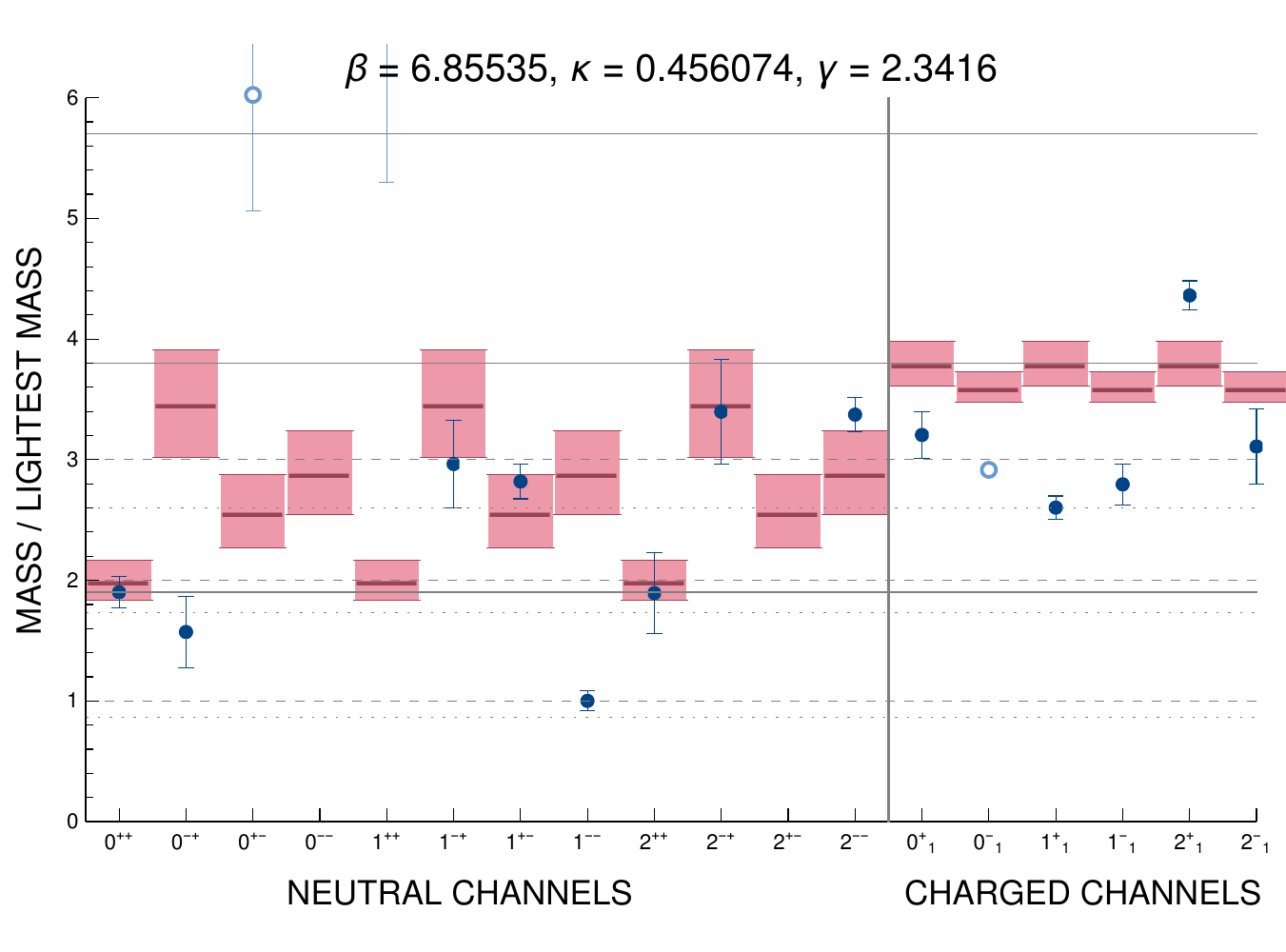}
    \vspace*{-.4cm}
    \caption{With BEH effect}
      \label{fig:sub:b}
    \end{subfigure}
    \\\vspace*{1cm}
    \hspace*{-.8cm}
    \begin{subfigure}{.52\textwidth}
    \centering
    \includegraphics[height=5.75cm]{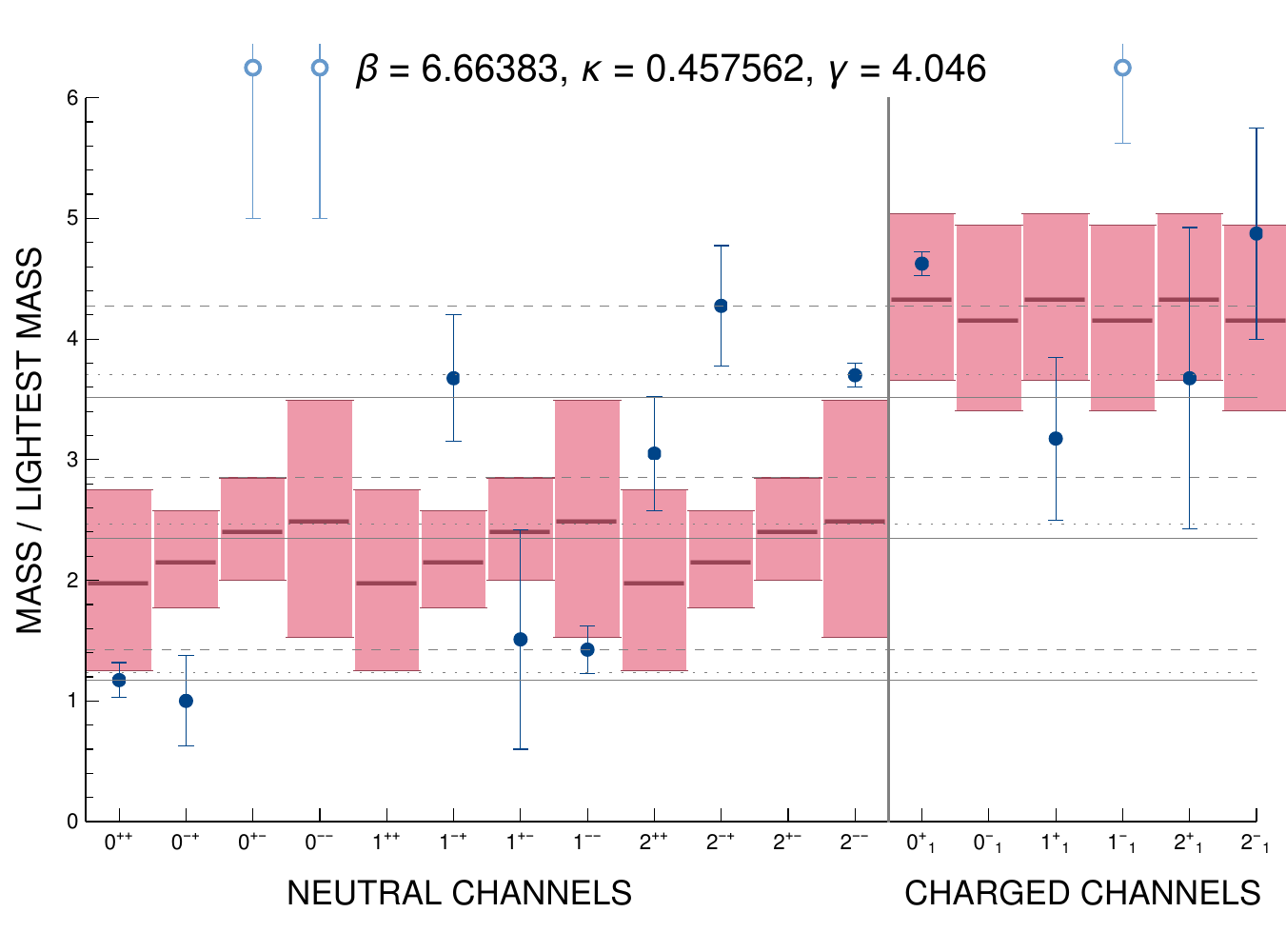}
    \vspace*{-.4cm}
    \caption{\centering Close to the transition between QCD-like\newline and BEH-like behavior}
      \label{fig:sub:QHTrans}
    \end{subfigure}
    \begin{subfigure}{.5\textwidth}
    \centering
    \includegraphics[trim= 15 0 0 0,clip,height=5.75cm]{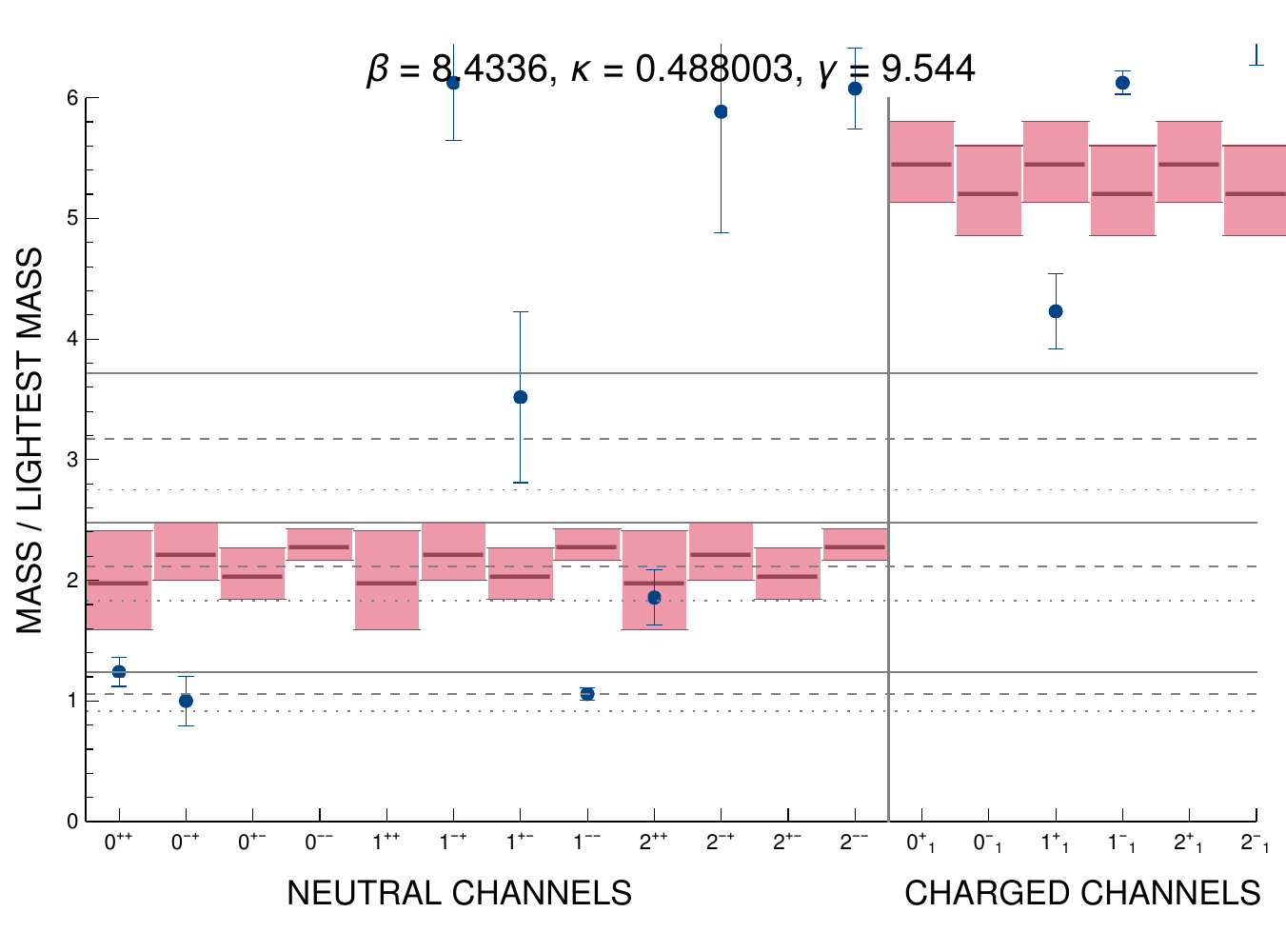}
    \caption{With BEH effect; note the similarities with \cref{fig:sub:b}
    % at larger $\beta/\kappa$ and $\gamma/kappa$
    }
      \label{fig:HLalt}
    \end{subfigure}
    \caption{
    Spectra, rescaled to the lightest state and extrapolated to the infinite-volume limit, for simulations at different points in the phase diagram.
    Error bars are statistical only, and empty circles indicate that only an upper bound on the ground-state could be obtained, and do not represent actual states.
    The grey horizontal lines give (multiples) of the basic scales of the FMS mechanism, extracted from the uncharged scalar and vector masses \cite{Maas:2018xxu}.
    These correspond to the Higgs mass (line), the heaviest gauge boson mass (dashed line) and the lighter gauge boson mass (dotted line) (being $\sqrt{3/4}$ of the heavier mass at tree-level). The relevant elastic thresholds in each channel (based on the stable ground-states) are shown as red blocks.
    Results \ref{fig:sub:a} and \ref{fig:sub:b} agree with the previous findings of \cite{Maas:2018xxu}.
    }
    \label{fig:spectrum}
\end{figure}

The spectra for a few different parameter sets are plotted in \cref{fig:spectrum}, and are consistent with previous results \cite{Maas:2018xxu,Maas:2016ngo} in available channels. 
Where no ground state could be determined, an empty circle in lighter blue indicates the estimate of the corresponding upper bound, and the values have been extrapolated to the continuum limit.
A striking feature is that we always observe a mass gap, in stark contrast to the ungapped elementary spectrum. So far, previous investigations \cite{Maas:2018xxu} suggest that this is not caused by nonperturbative, strong interactions in the unbroken $\SU(2)$ group. Ungapped physical spectra are, however, also possible and have been observed in other theories with a BEH effect but otherwise the same obstructions as for the present one \cite{Afferrante:2020hqe}.

It is to be expected that only the elementary mass scales, the mass of the Higgs and the two masses in the gauge boson spectrum, determine the masses of the physical spectrum. These have been included in \cref{fig:spectrum}. In the `Higgs-like' case it indeed appears that states cluster around multiples of these scales, confirming the prediction. The other is that the charged states cannot be mapped to an elementary state. Thus, the only explanation for the lightest state is that it is a non-trivial bound state. It was attempted to understand it via a leading-order constituent model \cite{Maas:2017xzh,Maas:2018xxu}. Lattice data seemed to agree with this at first, though with larger errors \cite{Maas:2018xxu}. The data shown in \cref{fig:spectrum}, which are more precise, seem to be less in favor. This also requires further understanding. In particular, it remains to be seen whether augmented perturbation theory beyond the leading order can determine their masses, or whether these states are genuinely nonperturbative.

Further, we confirm previous results in other theories \cite{Maas:2017wzi,Afferrante:2020hqe} that in the `BEH-like' case the lightest state is a vector, while otherwise it is a scalar particle. In the elementary spectrum, this is not necessary. Close to the transition, both states start to degenerate, as is also visible in \cref{fig:sub:QHTrans}. This was also observed in the $\SU(2)$ case \cite{Maas:2014pba,Evertz:1985fc}. In fact, all investigations so far yielded the same pattern \cite{Maas:2017wzi}.

Besides this, we find another uncharged stable state in both cases in the $0^{-+}$ channel. Within errors, it is found to be degenerate with the expected \cite{Maas:2017xzh} $0^{++}$ channel. This requires further investigation, as this state had not been studied before, and the presence of a bound state here is unexpected.

Furthermore, although in the `QCD-like' case the tensor state $2^{++}$ is stable, we note that it is much heavier than would be expected from $\SU(3)$ Yang--Mills theory in comparison to the (stable) $0^{-+}$ and $1^{--}$ states. Thus, the couplings are such that the scalar dynamics is not suppressed. In contrast, in the `BEH-like' case we do not observe any further uncharged stable states, which are statistically unambiguously below the elastic threshold, in agreement with \cite{Maas:2017xzh,Sondenheimer:2019idq}.

The charged sector is very interesting. In both cases we observe that charged states are generically heavier than uncharged ones, as would be naively expected in a constituent model based on equations \ref{eq:om} and \ref{eq:ob} \cite{Maas:2017xzh}. 
In the `QCD-like' case, most of the charged states are heavy, and it is not yet statistically significant, whether the lightest state is a (charged) scalar or a vector. 
This is different in the `BEH-like' case, where the vector states are lighter. Furthermore, there are two vector states of opposite parity, which are degenerate within statistical uncertainty. In addition, both in the scalar and the tensor channel states appear which seem to be bound. All of these states carry three times the custodial charge of the elementary scalar, and would not have been expected in a tree-level analysis. Indeed, even beyond the elementary spectrum only the vector state has been anticipated and previously observed \cite{Maas:2016ngo,Maas:2017xzh,Maas:2018xxu}. For the other states not yet an understanding exists, and requires further investigations. Such a rich charged spectrum would have not been anticipated based on usually analyses of GUTs \cite{Langacker:1980js}, and promises interesting phenomenological consequences, even beyond the scope of GUTs.

A comparison to an entirely different point in the parameter space in the 'BEH-like' region in \cref{fig:HLalt} shows a qualitatively very similar spectrum. In particular, the low-lying stable states have the same ordering, though quantitatively there are differences. However, some of the low-lying states in the charged sector may actually be scattering states, and a definitive statement requires further analysis. Nonetheless, that the qualitative features of the spectrum are essentially identical across the Higgs region has already been observed in other theories \cite{Maas:2014pba,Evertz:1985fc,Maas:2017wzi}. This seems to be a generic feature and is in stark contrast to QCD(-like) theories, where the quark mass alters the low-lying spectrum qualitatively.

Finally, the FMS mechanism cannot explain the observed pattern of the vector particle always being lighter than the scalar one, seen in \cref{fig:spectrum}, as well as in other theories and investigations \cite{Maas:2018xxu,Maas:2014pba,Evertz:1985fc,Maas:2017wzi}. This follows as the FMS mechanism merely describes the mapping between bound-state and elementary masses. Given that pushing the mass of the scalar lower leads to a transition into a QCD-like state, it seems to indicate that the dynamics of a too-light scalar makes the formation of a condensate impossible, in pronounced difference to perturbation theory \cite{Maas:2017wzi}. This is not yet understood, nor whether in regions of parameter space not accessible to lattice simulations this changes.

\section{Conclusions}
We have shown that, for a simple $\SU(3)$ Yang--Mills--Higgs model, the nonperturbative spectra is in qualitative disagreement with naive perturbative predictions, even at weak coupling. 
This is consistent with theoretical expectations \cite{Maas:2017xzh} and previous exploratory lattice simulations \cite{Maas:2016ngo,Maas:2018xxu}.
Our results therefore strongly imply that observables in a GUT-like theory cannot be adequately understood in terms of elementary fields.
It appears, however, that this problem can be mitigated by enforcing explicit gauge invariance, which allows composite-state effects, classified as `nonperturbative' in the standard approach, to still be treated within a perturbative framework. At least the studied uncharged states seem to be in reasonable agreement with such FMS-augmented perturbative predictions, but other channels seem to hold further challenges.

\subsection{Phenomenological implications}
The implications of our findings are twofold. 
The first is somewhat unsurprising: that bound-state effects resulting from gauge invariance can qualitatively change the spectrum for BSM models, rendering naive perturbative assumptions unreliable even at weak couplings.
The second is that such effects (usually classified as `nonperturbative' in standard perturbation theory) can still be treated via perturbative methods after all, provided we use an explicitly gauge invariant framework such as the FMS construction.

Unfortunately, these results give a somewhat pessimistic outlook for BSM model-building.
If the FMS approach is valid for understanding GUT theories, then this implies that finding an enlarged group with the correct low-energy spectrum is a much harder task.\footnote{See \cite{Sondenheimer:2019idq,Maas:2017wzi} for a more detailed review.}
However, our current results' consistency with the FMS approach gives some hope that this method could be in principle generalised to investigate properties of other GUT models which are less feasible to simulate on a lattice than our toy example.
\footnote{\mbox{It is reasonable to assume that if FMS gives valid predictions for $\SU(3)$, it should also do so for general $\SU(N)$ \cite{Maas:2017wzi,Maas:2017xzh,Sondenheimer:2019idq}.}}

A key point for model-building is the importance of systematic control: unless gauge invariance is enforced explicitly, then perturbative predictions can fail, even at weak coupling. 
Depending on the gauge group structure, large systematic errors can occur when gauge invariance is neglected, as seen, for example, with the appearance of additional custodially-charged states which are absent from the standard perturbative spectrum.

A better understanding of how bound-states are mapped between the broken and unbroken gauge groups when gauge-fixing is also important for understanding how effective gauge groups can be embedded into larger ones.

\subsection{Outlook}
The puzzle of the light pseudoscalar is still not understood, and will be addressed, along with a detailed analysis of the $\U(1)$-charged states, in a future article \cite{Dobson:2023}.
A parallel ongoing project is to repeat the spectroscopy for a similar model with the scalar in the adjoint representation. 
The adjoint case is phenomenologically interesting due to the multiple possible breaking patterns of the gauge group \cite{Dobson:2022ngz}, but is computationally much more expensive due to the presence of massless modes and the need for finite-momentum states and improved discretisation of operators \cite{Maas:2017xzh,Afferrante:2020hqe}.
In the adjoint case, the operators no longer map as cleanly onto elementary states, instead corresponding to more complex Wilson loops and glueballs with insertions of scalars at different points.\footnote{I.e., the equivalent of (\ref{eq:om},\ref{eq:ob}) would be of the form \cite{Dobson:2023} \begin{equation*}
    \omega_{\mu_1,\dots,\mu_N;\{r_1,\dots,r_Q\}} = \tr\bq*{
    (D_{\mu_N}\dots D_{r_Q+1}\phi)
    \dots
    (D_{\mu_{r_2}}\dots D_{\mu_{r_1+1}}\phi)
    (D_{\mu_{r_1}}\dots D_{\mu_1}\phi)
    }.
    \end{equation*}
    }
Such operators are expected to be even noisier than those for the fundamental case.

A longer term aim, for both the fundamental and adjoint cases, is to use a more detailed picture of the phase diagram to understand whether the FMS construction is valid throughout the 'BEH-like' phase, or if genuinely nonperturbative effects arise in some region. 

\acknowledgments
\noindent We are grateful to René Sondenheimer, Pascal Törek, Vincenzo Afferrante and Simon Plätzer for helpful discussions.
This project used the Vienna Scientific Cluster (VSC). ED and BR have been supported by the Austrian Science Fund FWF, grant P32760.

\printbibliography
\end{document}